\documentclass[authoryear,12pt,letter,3p]{jowarticle}
\usepackage{graphicx}
\usepackage{amsmath}
\usepackage{amssymb}
\usepackage{setspace}
\usepackage[all]{xy}
\usepackage{amsthm}
\usepackage{appendix}
\usepackage{csquotes}
\usepackage{longtable}
\usepackage{tabu}
\makeatletter
\renewcommand\section{\@startsection{section}{1}{\z@}{-3.25ex plus -1ex minus -.2ex}{1.5ex plus .2ex}{\normalsize\bf}}
\renewcommand\subsection{\@startsection{subsection}{2}{\z@}{-3.25ex plus -1ex minus -.2ex}{1.5ex plus .2ex}{\normalsize\bf}}
\renewcommand\subsubsection{\@startsection{subsubsection}{3}{\z@}{-3.25ex plus -1ex minus -.2ex}{1.5ex plus .2ex}{\normalsize\bf}}
\makeatother

\newtheorem{thm}{Theorem}

\newtheorem{prop}[thm]{Proposition}

\usepackage{multirow}

\begin{document}

\begin{frontmatter}
\title{Understanding Gauge}
\author{James Owen Weatherall}\ead{weatherj@uci.edu}
\address{Department of Logic and Philosophy of Science\\ University of California, Irvine, CA 92697}
\begin{abstract}I consider two usages of the expression ``gauge theory''.  On one, a gauge theory is a theory with excess structure; on the other, a gauge theory is any theory appropriately related to classical electromagnetism.  I make precise one sense in which one formulation of electromagnetism, the paradigmatic gauge theory on both usages, may be understood to have excess structure, and then argue that gauge theories on the second usage, including Yang-Mills theory and general relativity, do not generally have excess structure in this sense.
\end{abstract}
\end{frontmatter}

\doublespacing

\section{Introduction}

The word ``gauge'' is ubiquitous in modern physics.  Our best physical theories are described, in various contexts, as ``gauge theories''.  The ``gauge argument'' allegedly reveals the underlying ``logic of nature'' \citep{Martin}.  Our theories regularly exhibit ``gauge freedom'', ``gauge structure'', and ``gauge dependence''.  Unfortunately, however, it is far from clear that the term has some univocal meaning across the many contexts in which it appears.  It is a bit like ``liberal'' in American political discourse: it shows up everywhere, and no one knows what it means.

Here I will focus on two strands of usage.\footnote{There are still others---for instance, see \citet{Weyl}.  But I will not attempt a taxonomy.}  On the first strand, a ``gauge theory'' is a theory that exhibits excess structure or, in John Earman's words, ``descriptive fluff'' \citep{Earman}.\footnote{This way of speaking is common: see \citet{Rovelli}, \citet{Ismael+vanFraassen}, \citet{Redhead}, and many others.}  On this way of thinking about gauge, there is a mismatch between the mathematical structure used in the theory and the structure we take the world to have, in such a way that (perhaps) one could remove some structure from the theory without affecting its descriptive or representational power.  Most famously, \citet{Earman+Norton} argue that the so-called ``hole argument'' shows that general relativity is a gauge theory in just this sense; some have taken this as motivation for moving to a different, perhaps undiscovered, formalism for representing spacetime \citep{EarmanWEST}.\footnote{For an argument against this way of thinking about relativity theory, see \citet{WeatherallHoleArg}.  For more on the relationship between the standard formalism of relativity theory and at least one widely discussed alternative, see \citet{Rosenstock+Barrett+Weatherall}.}  Similar considerations have motivated some views on Yang-Mills theory \citep{Healey,Rosenstock+Weatherall1}.

The second strand of usage concerns a specific class of theories.  Here one uses ``gauge theory'' to refer to various generalizations of classical electromagnetism that share a certain mathematical structure.\footnote{Though I will say somewhat more about principal bundles and principal connections below, this will not be the occasion to review this formalism.  For more on this topic, see \citet{Bleecker}, \citet{Palais}, and \citet{WeatherallYM}.}  For instance, \citet{Trautman} defines ``gauge theory'' as follows:
\begin{quote}\singlespacing
For me, a gauge theory is any physical theory of a dynamic variable which, at the classical level, may be identified with a connection on a principal bundle. \citep[p. 26]{Trautman}
\end{quote}
This turns out to be a large class containing most of our fundamental theories, including all Yang-Mills theories, general relativity, and Newton-Cartan theory.  It is in this context that physicists seem to speak most often of gauge theories, usually as a synonym for ``Yang-Mills theories''.

It is easy to imagine that the two strands are closely related, and in particular, that all gauge theories in the second sense are also gauge theories in the first sense.  But as I will argue below, this is a mistake.  In particular, I will articulate a precise sense in which electromagnetism---the paradigmatic example on both strands---may be understood to have excess structure, and thus to be a gauge theory in the first sense.  I will then consider whether other theories, such as Yang-Mills theory and general relativity, have excess structure in the same sense.  I will argue that they do not.  It follows that on at least one precise sense of what it means for a theory to have excess structure, the two strands of usage described above come apart.%\footnote{The present paper is part of a larger project; an overview is given in \citet{WeatherallLandry}.}

\section{Two approaches to electromagnetism}\label{sec:EM}

In what follows, we will consider electromagnetism on the fixed background of Minkowski spacetime, $(M,\eta_{ab})$.\footnote{Minkowski spacetime is a relativistic spacetime where $M$ is $\mathbb{R}^4$ and $\eta_{ab}$ is flat and complete.  We focus on this case for convenience; little of consequence turns on the limitation.  Note that we work in the abstract index notation, following the sign conventions of \citet{MalamentGR}.} In this setting, there are two ways of characterizing models of ordinary electromagnetism.\footnote{These two ways of thinking about electromagnetism are described in somewhat more detail in \citet{WeatherallTE}.  Of course, there are also other ways of characterizing the models of electromagnetism---including using the principal bundle formalism discussed below.}

On one characterization, the principal dynamical variable is the electromagnetic field, represented by a two form $F_{ab}$ on spacetime.  The electromagnetic field is required to satisfy Maxwell's equations, which may be expressed as $\nabla_{[a}F_{bc]}=\mathbf{0}$ and $\nabla_a F^{ab} = J^b$, where $\nabla$ is the Minkowski derivative operator and $J^a$ is a smooth vector field representing the charge-current density on spacetime.  A model of the theory on this characterization might be written as a triple $(M,\eta_{ab},F_{ab})$, where $F_{ab}$ is any closed two form (i.e., any two form satisfying the first of Maxwell's equations).\footnote{\label{provisos} One might also stipulate a source term, $J^b$, along with the model, or perhaps limit attention to models for which the source term satisfies certain ``physically reasonable'' conditions.  Here it suffices to permit arbitrary sources, and to ``read off'' the charge-current density from the divergence of $F_{ab}$, using the second of Maxwell's equations.}  Call this formulation of the theory EM$_1$.

On the second characterization, the dynamical field is the four-vector potential, represented by a one form $A_a$ on spacetime.  This field is required to satisfy a single differential equation: $\nabla_a\nabla^a A^b - \nabla^b\nabla_a A^a = J^b$, where again $J^a$ is the charge-current density.  A model of the theory may again be represented by a triple, $(M,\eta_{ab},A_a)$, where $A_a$ is any one form.\footnote{The same remarks apply here as in fn. \ref{provisos}.} I will call this formulation of the theory EM$_2$.

These two formulations are systematically related.  Given any model $(M,\eta_{ab},A_a)$ of EM$_2$, I can always define an electromagnetic field by $F_{ab}=\nabla_{[a}A_{b]}$.  Since any $F_{ab}$ thus defined is exact, it must also be closed, and thus the resulting triple $(M,\eta_{ab},F_{ab})$ is a model of EM$_1$; moreover, this $F_{ab}$ is associated with the same charge-current density as $A_a$.  Conversely, given any model $(M,\eta_{ab},F_{ab})$ of EM$_1$, since $F_{ab}$ is closed, it must also be exact, and thus there exists a one form $A_a$ such that $F_{ab}=\nabla_{[a}A_{b]}$.\footnote{This result holds globally because $M$ is homeomorphic to $\mathbb{R}^4$.  In the case of more general spacetimes, one would have to work locally; this difference raises interesting issues, but they are not relevant to the current discussion.}  The triple $(M,\eta_{ab},A_b)$ is then a model of EM$_2$, again with the same charge-current density.

However, there is an important asymmetry in this relationship.  Given any model of EM$_2$, there exists a \emph{unique} corresponding model of EM$_1$, because any smooth one form has a unique exterior derivative.  But the converse is not true: given a model $(M,\eta_{ab},F_{ab})$ of EM$_1$, there will generally be \emph{many} corresponding models of EM$_2$, since if $A_a$ is such that $F_{ab}=\nabla_{[a}A_{b]}$, then $A'_a = A_a + \nabla_a \chi$, for any smooth scalar field $\chi$, \emph{also} satisfies $F_{ab} = \nabla_{[a}A'_{b]} = \nabla_{[a}A_{b]} + \nabla_{[a}\nabla_{b]} \chi$, because for any smooth scalar field, $\nabla_{[a}\nabla_{b]}\chi=\mathbf{0}$.  Transformations $A_a\mapsto A'_a$ of this form are sometimes known as \emph{gauge transformations}.

It is this asymmetry that, I claim, supports the common view that electromagnetism has excess structure.  The idea is that EM$_1$ and EM$_2$ both have all of the resources necessary to represent classical electromagnetic phenomena.  Indeed, in both cases, one may take the empirical content of electromagnetism to be fully exhausted by the electromagnetic field associated with a given model---either directly in the case of EM$_1$, or as derived above in the case of EM$_2$.  But there are \emph{prima facie} distinct models of EM$_2$ associated with the same electromagnetic field.  Thus it would seem that these models of EM$_2$, though they differ in their mathematical properties, should be taken to have the same representational capacities.  Intuitively, then, whatever structure distinguishes models of EM$_2$ related by gauge transformations must be irrelevant to the representational capacities of the models, at least as long as the empirical content is exhausted by the associated electromagnetic field.  It is in this sense that electromagnetism---or really, EM$_2$---has excess structure.

\section{Comparing structure}

In the next section, I will make the intuitive argument just given precise.  First, however, I will take a detour through pure mathematics.  Mathematical objects often differ in how much structure they have.  For instance, topological spaces have more structure than sets: a topological space $(X,\tau)$ consists in a set $X$, along with something more, namely a collection $\tau$ of open subsets of $X$ satisfying certain properties.  Similarly, an inner product space has more structure than a vector space, and a Lie group has more structure than a smooth manifold.  In the present section, I will use some basic category theory to capture these judgments as mathematical relationships between the theories.\footnote{For background on basic category theory, see \citet{MacLane} or \citet{Leinster}; due to space constraints, I take notions like ``category'' and ``functor'' for granted.}

%In what follows, a \emph{category} $\mathcal{C}$ consists in a collection of objects $A,B,C,\ldots$, a collection of arrows $f,g,h,\ldots$, and a pair of maps $dom$ and $cod$ (read ``domain'' and ``codomain'') each taking arrows to objects.  (We will write $f:A\rightarrow B$ to indicate that, for some arrow $f$, $dom(f)=A$ and $cod(f)=B$.)  In addition, for any arrows $f,g$ such that $cod{f}=dom{g}$, there exists an arrow $g\circ f : dom{f}\rightarrow cod{g}$ called the \emph{composition of $f$ and $g$}, and for any object $A$, there exists an arrow $1_A: A\rightarrow A$ called the \emph{identity arrow}.  Together, these must satisfy the following properties: (1) for any arrows $f,g,h$, $(h\circ g)\circ f = h\circ(g\circ f)$; and (2) for any arrow $f:A\rightarrow B$, $f\circ 1_A = f = 1_B\circ f$.  A \emph{functor} $F$ between categories $\mathcal{C}$ and $\mathcal{D}$, written $F:\mathcal{C}\rightarrow\mathcal{D}$, consists in maps from objects of $\mathcal{C}$ to objects of $\mathcal{D}$ and from arrows of $\mathcal{C}$ to arrows of $\mathcal{D}$, with the properties that (1) for any arrow $f:A\rightarrow B$ of $\mathcal{C}$, $F(f):F(A)\rightarrow F(B)$, (2) for any object $A$ of $\mathcal{C}$, $F(1_A)=1_{F(A)}$, and (3) for any arrows $f,g$ of $\mathcal{C}$, $F(f\circ g)=F(f)\circ F(g)$.

To begin, recall that various mathematical theories may be associated with categories.  For instance, there is a category $\mathbf{Set}$, whose objects are sets and whose arrows are functions.  There is a category $\mathbf{Top}$, whose objects are topological spaces and whose arrows are continuous functions.  There are also functors between such categories.  For instance, there is a functor $F:\mathbf{Top}\rightarrow\mathbf{Set}$ that takes every topological space $(X,\tau)$ to the set $X$, and takes every continuous map $f:(X,\tau)\rightarrow (X',\tau')$ to the function $f:X\rightarrow X'$.  Functors of this sort are often called ``forgetful'', because, intuitively speaking, they take objects of one category and forget something about them: in this case, they take topological spaces $(X,\tau)$ and forget about $\tau$.

How can one tell whether a given functor is forgetful?  There is a simple but insightful theory available, due to \citet{Baez+etal}.\footnote{See also \citet{BarrettCounting}.}  It requires a few further definitions, concerning properties that a functor $F:\mathcal{C}\rightarrow\mathcal{D}$ may have.  First, we will say that $F$ is \emph{full} if for all objects $A,B$ of $\mathcal{C}$, the map $(f:A\rightarrow B)\mapsto (F(f):F(A)\rightarrow F(B))$ induced by $F$ is surjective.  Similarly, $F$ is \emph{faithful} if for all pairs of objects in $\mathcal{C}$, the induced map on arrows is injective.  And $F$ is \emph{essentially surjective} if for every object $X$ of $\mathcal{D}$, there is some object $A$ of $\mathcal{C}$ and arrows $f:F(A)\rightarrow X$ and $f^{-1}:X\rightarrow F(A)$ such that $f^{-1}\circ f=1_{F(A)}$ and $f\circ f^{-1}=1_X$.  (Such an arrow $f$ is an \emph{isomorphism}, so essentially surjective functors are surjective on objects ``up to isomorphism''.)

If a functor $F:\mathcal{C}\rightarrow \mathcal{D}$ is full, faithful, and essentially surjective, then the functor is said to realize an \emph{equivalence} of categories.  In such cases, $F$ forgets \emph{nothing}.  Otherwise a functor is forgetful.  In particular, a functor forgets (only) \emph{structure} if it is faithful and essentially surjective, but not full.  A functor forgets (only) \emph{properties} if it is full and faithful, but not essentially surjective.  And a functor forgets (only) \emph{stuff} if it is full and essentially surjective, but not faithful.  In general, a given functor may be forgetful in more than one of these ways, but not in any \emph{other} ways: any functor may be written as the composition of three functors that forget (no more than) structure, properties, and stuff, respectively.

The best way to make this categorization plausible is by considering examples.  For instance, the functor we have already considered, $F:\mathbf{Top}\rightarrow\mathbf{Set}$, forgets only structure.  This is because every set corresponds to some topological space (or other), which means that $F$ is essentially surjective.  Similarly, any distinct continuous functions $f,f':(X,\tau)\rightarrow(X',\tau')$ must be distinct as functions, so $F$ is faithful.  But $F$ is not full, because not every function $f:X\rightarrow X'$ is continuous, given topologies on $X$ and $X'$.  So in this case, the classification captures the pre-theoretic intuition with which we began.

Similarly, we can define categories $\mathbf{Grp}$ and $\mathbf{AbGrp}$, whose objects are groups and Abelian groups, respectively, and whose arrows are group homomorphisms; then there is a functor $G:\mathbf{AbGrp}\rightarrow\mathbf{Grp}$ that takes Abelian groups and group homomorphisms to themselves.  This functor is full and faithful, since it just acts as the identity on group homomorphisms between Abelian groups.  But it is not essentially surjective, because not every group is Abelian.  So this functor forgets only properties---namely, the property of being Abelian.  And finally, we can a define a functor $H:\mathbf{Set}\rightarrow 1$, where $1$ is the category with one object and one arrow (the identity on the one object).  This functor takes every set to the unique object of $1$, and every arrow to the unique arrow of $1$.  It is clearly full and essentially surjective, but not faithful, so it forgets only stuff.  To see how, note that we may think of $1$ as the category with the empty set as its only object; thus $H$ forgets all of the elements of the sets.

This classification of functors gives us a criterion for when a mathematical theory $T_1$ may be said to have more structure than another theory $T_2$: namely, when there exists a functor from the category associated with $T_1$ to the category associated with $T_2$ that forgets structure.  Given two categories, there may be multiple functors between them, and it may be that not all such functors forget structure, even if there exists one that does.  This means that comparative judgments of ``amount of structure'' between theories should be understood as relative to a choice of functor.  This flexibility is a virtue: it allows us to explore various ways in which theories may be related.

\section{A Diagnostic Tool}

I will now return to the question of interest.  To begin, I will use the criterion just developed to make the intuitive argument at the end of section \ref{sec:EM} precise.  I will define two categories, corresponding to the two formulations of electromagnetism already discussed, and then we define a functor between them that captures the relationship already discussed between EM$_1$ and EM$_2$.

The first category, $\mathbf{EM}_1$, has models $(M,\eta_{ab},F_{ab})$ of EM$_1$ as objects, and as arrows has maps that suitably preserve this structure.  For present purposes, we will take these to be isometries of Minkowski spacetime that preserve the electromagnetic field, so that given two models, $(M,\eta_{ab},F_{ab})$ and $(M,\eta_{ab},F'_{ab})$, an arrow $\chi:(M,\eta_{ab},F_{ab})\rightarrow (M,\eta_{ab},F'_{ab})$ will be an isometry of $(M,\eta_{ab})$ such that $\chi^*(F'_{ab})=F_{ab}$.  Likewise, we may define a category $\mathbf{EM}_2$ whose objects are models $(M,\eta_{ab},A_a)$ of EM$_2$ and whose arrows are isometries of Minkowski spacetime that preserve the 4-vector potential. Given these categories, the map defined above, taking models $(M,\eta_{ab},A_a)$ of EM$_2$ to models $(M,\eta_{ab},\nabla_{[a}A_{b]})$ of EM$_1$, becomes a functor $F:\mathbf{EM}_2\rightarrow\mathbf{EM}_1$ that take arrows of $\mathbf{EM}_2$---which, recall, are isometries of Minkowski spacetime with an additional property---to the arrow of $\mathbf{EM}_1$ corresponding to the same isometry.  (This action on arrows is well-defined because, given any arrow $\chi:(M,\eta_{ab},A_a)\rightarrow (M,\eta_{ab},A'_a)$ of $\mathbf{EM}_2$, $\nabla_{[a}A_{b]} = \nabla_{[a}(\chi^*(A'_{b]}))=\chi^*(\nabla_{[a}A'_{b]})$.\footnote{See \citet{WeatherallTE} for further details on this functor.})  We then have the following result.

\begin{prop}\label{forgetProp} $F$ forgets only structure.\end{prop}
Proof. $F$ is clearly faithful and essentially surjective. To see that it is not full, consider the object $A=(M,\eta_{ab},\mathbf{0})$ of $\mathbf{EM}_1$.  The object $X=(M,\eta_{ab},\mathbf{0})$ of $\mathbf{EM}_2$ maps to $A$.  But now consider any non-constant scalar field $\psi$.  Then $X'=(M,\eta_{ab},\nabla_a\psi)$ is a model of $\mathbf{EM}_2$ and $F(X)=F(X')=A$. If $F$ were full, then, there would have to be an arrow $f:X\rightarrow X'$ such that $F(f)=1_A$, but this is impossible, since by construction, there are \emph{no} arrows between $X$ and $X'$. \hspace{.25in}$\square$

Prop. \ref{forgetProp} provides a precise sense in which EM$_2$ has more structure than EM$_1$: the functor realizing the natural relationship between the theories forgets (only) structure. Recall that the intuitive argument was that there are distinct models of EM$_2$ corresponding to a single model of EM$_1$, and thus there must be features of the models of EM$_2$ that distinguish them, without making any difference to their empirical content.  The present argument, meanwhile, is that there are models of EM$_2$ that fail to be isomorphic---by the standard of isomorphism used in defining EM$_2$---even though the corresponding models of EM$_1$ are isomorphic, or even identical.  This is captured in the formalism by the fact that there are arrows in $\mathbf{EM}_1$, which we may interpret as ``structure preserving maps'' between models of EM$_1$, that are not structure preserving maps between models of EM$_2$.  The structure that these maps do not preserve is the structure that, on the intuitive argument, distinguished models of EM$_2$.

I take this to be strong evidence that the formal criterion given by forgetful functors captures the sense in which electromagnetism has excess structure. And since electromagnetism is the paradigmatic example of a gauge theory, I take this to be the sense of ``excess structure'' associated with the first strand of usage.\footnote{There are other examples of theories with excess structure in this sense, too---for instance, Newtonian gravitation, which is also sometimes described as a gauge theory \citep[p. 248]{MalamentGR}, may be understood to have excess structure in just this sense. To see this, consider the discussion in \citet[\S 6]{WeatherallTE} and observe that the functor from $\mathbf{NG}_1$ to $\mathbf{GNG}$ given by the Trautman geometrization theorem is faithful and essentially surjective, but not full.}  Of course, there may be other senses in which a theory might be thought to have excess structure, but I will not consider that question further here.  Rather, I will stipulate that the criterion developed here is salient, and turn to a different question.  Do gauge theories in the second sense---that is, the theories Trautman identifies---have excess structure?

No. First, consider electromagnetism, formulated now as a theory whose dynamical variable is a connection on a principal bundle over Minowski spacetime---that is, electromagnetism formulated as a gauge theory in Trautman's sense.  Call this theory EM$_3$.  Models of EM$_3$ may be written $(P,\omega_{\alpha})$, where $P$ is the total space of the (unique, trivial) principal bundle $U(1)\rightarrow P\xrightarrow{\pi} M$ over Minkowski spacetime and $\omega_{\alpha}$ is a principal connection.\footnote{Again, see \citet{Palais}, \citet{Bleecker}, or \citet{WeatherallYM} for more details on this theory.}  This theory is closely related to both EM$_1$ and EM$_2$ as already discussed: given any (global) section $\sigma:M\rightarrow P$, we may define a 4-vector potential $A_a$ as the pullback along $\sigma$ of $\omega_{\alpha}$: $A_a=\sigma^*(\omega_{\alpha})$.  Similarly, we may define an electromagnetic field tensor $F_{ab}$ as the pullback along $\sigma$ of the \emph{curvature} of the connection, defined by $\Omega_{\alpha\beta}=d_{\alpha}\omega_{\beta}$, where $d$ is the exterior derivative on $P$: $F_{ab}=\sigma^*(\Omega_{\alpha\beta})$.  Thus $A_a$ and $F_{ab}$ may be thought of as representatives on $M$ of the connection and curvature on $P$.  In general, $A_a$ will depend on the choice of section $\sigma$, whereas $F_{ab}$ will not depend on that choice because $U(1)$ is an Abelian group.

Given this characterization of the theory, we can define yet another category, $\mathbf{EM}_3$, as follows: the objects of $\mathbf{EM}_3$ are models of EM$_3$, and the arrows are principal bundle isomorphisms $(\Psi,\psi)$ that preserve both the connection on $P$ and the metric on $M$: that is, pairs of diffeomorphisms $\Psi:P\rightarrow P$ and $\psi:M\rightarrow M$ such that $\psi^*(\eta_{ab})=\eta_{ab}$, $\Psi^*(\omega_{\alpha})=\omega_{\alpha}$, $\pi\circ\Psi = \psi\circ\pi$, and $\Psi(xg)=\Psi(x)g$ for any $x\in P$ and any $g\in U(1)$.  Then we may define a functor $\tilde{F}:\mathbf{EM}_3\rightarrow\mathbf{EM}_1$ as follows: $\tilde{F}$ act on objects as $(P,\omega_{\alpha})\mapsto (M,\eta_{ab},\sigma^*(\Omega_{\alpha\beta}))$, for any global section $\sigma:M\rightarrow P$, and $\tilde{F}$ acts on arrows as $(\Psi,\psi)\mapsto \psi$.  (Again, this action on arrows is well-defined. Choose any section $\sigma:M\rightarrow P$.  Then $\Psi^{-1}\circ\sigma\circ\psi$ is also a section of $P$.  But since $F_{ab}=\sigma^*(\Omega_{\alpha\beta})$ and $F'_{ab}=\sigma^*(\Omega'_{\alpha\beta})$ are independent of the choice of section, $F_{ab}=(\Psi^{-1}\circ\sigma\circ\psi)^*(\Omega_{\alpha\beta})=\psi^*\circ\sigma^*\circ\Psi^{-1}{}^*(\Omega_{\alpha\beta})=\psi^*\circ\sigma^*(\Omega'_{\alpha\beta})=\psi^*(F'_{ab})$.)
\begin{prop}$\tilde{F}$ forgets nothing.\label{PCforgets}\end{prop}
Proof. First consider an arbitrary object $(M,\eta_{ab},F_{ab})$ of $\mathbf{EM}_1$ and pick some $A_a$ such that $\nabla_{[a}A_{b]}=F_{ab}$.  Then given any section $\sigma:M\rightarrow P$ and any one-form $A_a$, we can define a connection $\omega_{\alpha}$ on $P$ by assigning to each point $x\in\sigma[M]$ the one form $\pi^*(A_a)$, and then extending this field on $\sigma[M]$ to all of $P$ by requiring it to be equivariant.  It follows that $\tilde{F}$ is essentially surjective.  Now consider any two objects $(P,\omega_{\alpha})$ and $(P,\omega'_{\alpha})$ of $\mathbf{EM}_3$, and suppose there exist arrows $(\Psi,\psi),(\Psi',\psi'):(P,\omega_{\alpha})\rightarrow (P,\omega'_{\alpha})$ such that $\tilde{F}((\Psi,\psi))=\psi=\psi'=\tilde{F}((\Psi',\psi'))$.  It follows that $(\Psi'\circ\Psi^{-1},1_M)$ is a vertical bundle automorphism such that $(\Psi'\circ\Psi^{-1})^*(\omega'_{\alpha})=\omega'_{\alpha}$.  But by the equivariance of $\omega'_{\alpha}$, this is only possible if $\Psi'\circ \Psi^{-1} = 1_P$.  Since $\Psi$ and $\Psi^{-1}$ are bijective, it follows that $\Psi=\Psi'$. Thus $\tilde{F}$ is faithful.  Finally, suppose there is an arrow $\psi:\tilde{F}((P,\omega_{\alpha}))\rightarrow\tilde{F}((P,\omega'_{\alpha}))$.  Then $\psi$ is an isometry of Minkowski spacetime such that, for any section $\sigma:M\rightarrow P$, $\psi^*(\sigma^*(d_{\alpha}\omega'_{\beta}))=\sigma^*(d_{\alpha}\omega_{\beta})$.  Fix $\sigma:M\rightarrow P$. It follows that there exists a section $\tilde{\sigma}:M\rightarrow P$ such that $\tilde{\sigma}^*(\omega_{\alpha})=\psi^*(\sigma^*(\omega'_{\alpha}))$.  We may then define a diffeomorphism $\Psi:P\rightarrow P$ as follows.  For any $x\in P$, there exists some $g\in U(1)$ such that $x=\tilde{\sigma}\circ\pi(x)g$.  This relationship determines a smooth map $g:P\rightarrow U(1)$ satisfying $g(xh)=g(x)h$ for any $h\in U(1)$.  We then define $\Psi$ by $x\mapsto \sigma\circ\psi\circ\pi(x)g(x)$ for all $x\in P$.  This map $\Psi:P\rightarrow P$ is a diffeomorphism such that $\Psi(xh)=\sigma\circ\psi\circ\pi(xh)g(xh)=\sigma\circ\psi\circ\pi(x)g(x)h=\Psi(x)h$ and $\pi\circ\Psi=\pi\circ\sigma\circ\psi\circ\pi=\psi\circ \pi$.  Thus $(\Psi,\psi):P\rightarrow P$ is a principal bundle isomorphism.  Moreover, since for any $x\in \tilde{\sigma}[M]$, $(\Psi_x)^*(\omega'_{\alpha})=(\pi_x)^*\circ (\psi_{\pi(x)})^*\circ \sigma_{\psi\circ\pi(x)}^*(\omega'_{\alpha})=(\pi_x)^*\circ(\tilde{\sigma}_{\pi(x)})^*(\omega_{\alpha})=(\omega_{\alpha})_{|x}$, it follows by the equivariance of $\omega_{\alpha}$ and $\omega'_{\alpha}$ that $\Psi^*(\omega'_{\alpha})=\omega_{\alpha}$.  So $\tilde{F}$ is full.\hspace{.25in}$\square$

This result shows that EM$_3$ does not have excess structure in the sense that EM$_2$ does.  To extend this to other gauge theories in the second sense, however, requires more work.  The reason is that the criterion we have been using requires us to have \emph{two} formulations, both of which are taken to be descriptively adequate and empirically equivalent, which we then compare.  In other cases of interest, though, such as non-Abelian Yang-Mills theory or general relativity, it is not clear that we have a plausible second theory to consider.

Still, there is something one can say.  It concerns the role of ``gauge transformations'' between models of EM$_2$, as described at the end of section \ref{sec:EM}. These are maps that relate models of EM$_2$ that have the same representational capacities even though they are not isomorphic.  The criterion of excess structure described here, meanwhile, requires the existence of a functor between categories of models that fails to be full---or in other words, a standard of comparison between the theories relative to which one formulation has ``more'' arrows than the other, or alternatively, relative to which one of the formulations is ``missing'' arrows.\footnote{This way of speaking may be made somewhat more precise, by showing how adding arrows corresponding to gauge transformations to $\mathbf{EM}_2$ can lead to a new category that is equivalent to $\mathbf{EM}_1$.  See \citet{WeatherallTE}.}  This suggests a rule of thumb for whether a theory, or a formulation of a theory, has excess structure in the sense described here: namely, if the theory has models that are not isomorphic, but which nevertheless we interpret as having precisely the same representational content.  Indeed, whereas the criterion discussed above tells us when on theory or formulation has more structure than another, this second criterion evaluates whether \emph{any} alternative formulation could have less structure and still do the same descriptive work---at least without equivocating between physical situations we now think are distinct.

How can we put this rule of thumb to work?  Suppose you are given a theory and a collection of maps taking models to models with the same representational capacities---that is, one is presented with a candidate ``gauge theory'' and a class of ``gauge transformations''.  One may then ask: are these gauge transformations naturally construed as isomorphisms of the models of the theory, understood as mathematical objects?  If the answer is ``yes'', then it would seem that these maps do not signal excess structure, since these maps would not be ``missing'' from a natural category of models; conversely, if the answer is ``no'', then there likely is excess structure in the formulation.

Applying this diagnostic to some examples of gauge theories in the second sense above, we immediately see that the moral concerning EM$_3$ generalizes to other Yang-Mills theories.  For instance, models of an arbitrary Yang-Mills theory with structure group $G$ may be written $(P,\omega^{\mathfrak{A}}{}_{\alpha})$, where $P$ is a principal $G$-bundle over some spacetime $(M,g_{ab})$ and $\omega^{\mathfrak{A}}{}_{\alpha}$ is a principal connection on $P$.\footnote{Again, for more on the notation used here, see \citet{WeatherallYM}.}  In this setting, a ``gauge transformation'' is often defined as a (vertical) principal bundle automorphisms $(\Psi,1_M)$ relating models $(P,\omega^{\mathfrak{A}})$ and $(P,\Psi_*(\omega^{\mathfrak{A}}{}_{\alpha}))$.\footnote{For instance, see \citet[\S 3.2]{Bleecker}.}  But these maps are just a special class of connection- and metric-preserving principal bundle isomorphisms, and so although they do map between models with the same representational resources, they are not ``extra'' maps, in the sense of the gauge transformations of EM$_2$.  So Yang-Mills theory does not have excess structure in the sense discussed here.

Likewise, for general relativity, we characterize models of the theory as relativistic spacetimes, $(M,g_{ab})$.  Here ``gauge transformations'' are often taken to be diffeomorphisms $\varphi:M\rightarrow M$ relating models $(M,g_{ab})$ and $(M,\varphi_*(g_{ab}))$.\footnote{For instance, see \citet{Wald} or \citet{Earman+Norton}.}  But once again, these maps are just isometries---i.e., they are just isomorphisms of Lorentzian manifolds.  So here, too, there is no excess structure.

\section{Conclusion}

I have isolated two strands of usage of the expression ``gauge theory'' in physics and philosophy of physics.  According to one, a gauge theory is a theory that has excess structure; according to the other, a gauge theory is any theory whose dynamical variable is a connection on a principal bundle.  I then endeavored to make precise the sense in which the paradigmatic example of a gauge theory (according to both strands)---classical electromagnetism---may be construed as having excess structure.  From this discussion, I extracted a general criterion for when a theory has excess structure.  From this criterion, I argued that gauge theories in the second sense need not have excess structure---and indeed, Yang-Mills theory and general relativity should not be construed as having excess structure in the sense that one formulation of electromagnetism does.

\section*{Acknowledgments}
This material is based upon work supported by the National Science Foundation under Grant No. 1331126.  Thank you to Thomas Barrett, Gordon Belot, Ben Feintzeig, Richard Healey, David Malament, Sarita Rosenstock, and David Wallace for helpful discussions related to the material in this paper, to Thomas Barrett, Ben Feintzeig, David Malament, JB Manchak, and Sarita Rosenstock for comments on an earlier draft, and to my fellow symposiasts---Thomas Barrett, Hans Halvorson, and Sahotra Sarkar---for a stimulating session at the 2014 PSA Biennial Meeting, at which this work was presented.

\singlespacing

\newpage
\section*{Erratum to ``Understanding Gauge''}

Prop. \ref{PCforgets} is false as stated in the published version of this article [Weatherall, J. O. (2016) ``Understanding Gauge'' \emph{Philosophy of Science} \textbf{83}(5), 1039--1049].  The functor $\tilde{F}$ defined on p. 1046 is full and essentially surjective, but it is not faithful.  The proposition should be amended to read:
\setcounter{thm}{1}
\renewcommand{\thethm}{\arabic{thm}$^\star$}%
\begin{prop}\label{erratumProp}
  $\tilde{F}$ forgets only stuff.
\end{prop}
\noindent The original proof of Prop. \ref{PCforgets} establishes Prop. \ref{erratumProp}; only the part of the proof that seeks to establish faithfulness is faulty.

It is a direct consequence of Prop. \ref{erratumProp} that $\tilde{F}$ does not forget structure, since it is full.  In this sense, the arguments of the paper following Prop. \ref{PCforgets} go through without modification: EM$_3$ does not have ``excess structure'', in the sense discussed in this paper, relative to EM$_1$.  On the other hand, the relationship reflected by Prop. \ref{erratumProp} indicates that these theories should not be viewed as equivalent, and suggests that in fact EM$_3$ has \emph{less} structure than EM$_1$.  I will defer further discussion of this situation to future work.

I discovered the error on my own, but I am grateful to James Nguyen, Nic Teh, and Laura Wells, whose paper ``Why surplus structure is not superfluous'' [forthcoming in \emph{British Journal for Philosophy of Science}] prompted me to revisit Prop. \ref{PCforgets}, and to Thomas Barrett, Jeremy Butterfield, and Sarita Rosenstock for helpful discussions.


\begin{thebibliography}{23}
\expandafter\ifx\csname natexlab\endcsname\relax\def\natexlab#1{#1}\fi
\expandafter\ifx\csname url\endcsname\relax
  \def\url#1{\texttt{#1}}\fi
\expandafter\ifx\csname urlprefix\endcsname\relax\def\urlprefix{URL }\fi

\bibitem[{Baez et~al.(2004)Baez, Bartel, and Dolan}]{Baez+etal}
Baez, J., Bartel, T., Dolan, J., 2004. Property, structure, and stuff,
  available at: http://math.ucr.edu/home/baez/qg-spring2004/discussion.html.

\bibitem[{Barrett(2013)}]{BarrettCounting}
Barrett, T., 2013. How to count structure, unpublished.

\bibitem[{Bleecker(1981)}]{Bleecker}
Bleecker, D., 1981. Gauge Theory and Variational Principles. Addison-Wesley,
  Reading, MA, reprinted by Dover Publications in 2005.

\bibitem[{Earman(1989)}]{EarmanWEST}
Earman, J., 1989. World Enough and Space-Time. The MIT Press, Boston.

\bibitem[{Earman(2004)}]{Earman}
Earman, J., 2004. Laws, symmetry, and symmetry breaking: Invariance,
  conservation principles, and objectivity. Philosophy of Science 71~(5),
  1227--41.

\bibitem[{Earman and Norton(1987)}]{Earman+Norton}
Earman, J., Norton, J., 1987. What price spacetime substantivalism? the hole
  story. The British Journal for the Philosophy of Science 38~(4), 515--525.

\bibitem[{Healey(2007)}]{Healey}
Healey, R., 2007. Gauging What's Real: The Conceptual Foundations of
  Contemporary Gauge Theories. Oxford University Press, New York.

\bibitem[{Ismael and van Fraassen(2003)}]{Ismael+vanFraassen}
Ismael, J., van Fraassen, B., 2003. Symmetry as a guide to superfluous
  structure. In: Brading, K., Castellani, E. (Eds.), Symmetries in Physics:
  Philosophical Reflections. Cambridge University Press, Cambridge, pp.
  371--92.

\bibitem[{Leinster(2014)}]{Leinster}
Leinster, T., 2014. Basic Category Theory. Cambridge University Press,
  Cambridge.

\bibitem[{{Mac Lane}(1998)}]{MacLane}
{Mac Lane}, S., 1998. Categories for the Working Mathematician, 2nd Edition.
  Springer, New York.

\bibitem[{Malament(2012)}]{MalamentGR}
Malament, D., 2012. Topics in the Foundations of General Relativity and
  Newtonian Gravitation Theory. University of Chicago Press, Chicago.

\bibitem[{Martin(2002)}]{Martin}
Martin, C.~A., 2002. Gauge principles, gauge arguments, and the logic of
  nature. Philosophy of Science 69~(S3), 221--34.

\bibitem[{Palais(1981)}]{Palais}
Palais, R.~S., 1981. The Geometrization of Physics. Institute of Mathematics,
  National Tsing Hua University, Hsinchu, Taiwan, available at
  http://vmm.math.uci.edu/.

\bibitem[{Redhead(2003)}]{Redhead}
Redhead, M., 2003. The interpretation of gauge symmetry. In: Brading, K.,
  Castellani, E. (Eds.), Symmetries in Physics: Philosophical Reflections.
  Cambridge University Press, Cambridge, pp. 124--139.

\bibitem[{Rosenstock et~al.(2015)Rosenstock, Barrett, and
  Weatherall}]{Rosenstock+Barrett+Weatherall}
Rosenstock, S., Barrett, T., Weatherall, J.~O., 2015. On einstein algebras and
  relativistic spacetimes, unpublished manuscript.

\bibitem[{Rosenstock and Weatherall(2015)}]{Rosenstock+Weatherall1}
Rosenstock, S., Weatherall, J.~O., 2015. A categorical equivalence between
  generalized holonomy maps on a connected manifold and principal connections
  on bundles over that manifold, arXiv:1504.02401 [math-ph].

\bibitem[{Rovelli(2013)}]{Rovelli}
Rovelli, C., 2013. Why gauge?, arXiv:1308.5599 [hep-th].

\bibitem[{Trautman(1980)}]{Trautman}
Trautman, A., 1980. Fiber bundles, gauge fields, and gravitation. In: Held, A.
  (Ed.), General Relativity and Gravitation. Plenum Press, New York, pp.
  287--308.

\bibitem[{Wald(1984)}]{Wald}
Wald, R., 1984. General Relativity. University of Chicago Press, Chicago.

\bibitem[{Weatherall(2014)}]{WeatherallYM}
Weatherall, J.~O., 2014. Fiber bundles, yang-mills theory, and general
  relativity, arXiv:1411.3281 [physics.hist-ph].

\bibitem[{Weatherall(2015{\natexlab{a}})}]{WeatherallTE}
Weatherall, J.~O., 2015{\natexlab{a}}. Are newtonian gravitation and
  geometrized newtonian gravitation theoretically equivalent?
  ErkenntnisForthcoming. arXiv:1411.5757 [physics.hist-ph].

\bibitem[{Weatherall(2015{\natexlab{b}})}]{WeatherallHoleArg}
Weatherall, J.~O., 2015{\natexlab{b}}. Regarding the `{H}ole {A}rgument'. The
  British Journal for Philosophy of ScienceForthcoming. arXiv:1412.0303
  [physics.hist-ph].

\bibitem[{Weyl(1952)}]{Weyl}
Weyl, H., 1952. Space-Time-Matter. Dover, Mineola, NY.

\end{thebibliography}
\end{document}